\documentstyle{article}
\title{\bf On tachyon and sub-quantum phantom cosmologies}
\author{ Pedro F. Gonz\'{a}lez-D\'{\i}az.\\
Colina de los Chopos, Instituto de Matem\'{a}ticas y F\'{\i}sica
Fundamental\\ Consejo Superior de Investigaciones Cient\'{\i}ficas\\
Serrano 121, 28006 Madrid, SPAIN\\ }
\date{June 12, 2004}
\begin{document}
\maketitle \large \setlength{\baselineskip}{0.5cm} \vspace{1cm}

\begin{abstract}
This paper deals with dark and phantom energy in the tachyon and
sub-quantum models for dark energy. We obtain that the simplest
condition for such a regime to occur in these scenarios is that
the scalar field be Wick rotated to imaginary values which
correspond to an axionic field classically. By introducing
analytical expressions for the scale factor or the Hubble
parameter that satisfy all constraint equations of the used models
we show that such models describe universes which develop a big
rip singularity in the finite future.
\end{abstract}

\section{Introduction}

The increasing dropping of detailed analysis leading to quite an
ample observationally acceptable parameter space beyond the
cosmological-constant barrier [1] is opening the really intriguing
possibility that the universe is currently dominated by what is
dubbed as phantom energy [2]. Phantom energy has rather weird
properties which include [3]: an energy density increasing with
time, naively unphysical superluminal speed of sound, violation of
the dominant energy condition which would eventually allow
existence of inflating wormholes and ringholes [4], and ultimately
emergence of a doomsday singularity in the finite future which is
known as the {\it big rip} [5].

The regime for phantom energy takes place for state equation
parameters $\omega=p/\rho<-1$ and has been shown to occur in all
current dark-energy models. However, whereas the big rip
singularity is allowed to happen in quintessence [5] and k-essence
[6] models, it is no longer present in models based on generalized
Chaplygin-gas equations of state [7,8] having the form
$P=-A/\rho^n$, with $A$ and $n$ being constants. No discussion has
been so far made nevertheless on the occurrence of phantom energy
and big rip in the other major contender model for dark energy:
the tachyon matter scenario of Padmanabhan {\it et al.} [9] or its
sub-quantum generalization [10]. Abramo and Finelli have in fact
used [11] a Born-Infeld Lagrangian with a power-law potential and
recovered a nice dark-energy behaviour, but did not considered the
negative kinetic terms which appear to characterize phantom
energy.

While defining a phantom energy regime in such scenarios appears
to be rather straightforward, it is quite more difficult to obtain
an associated expression for the scale factor of the accelerating
universe which allows us to see whether or not a big rip
singularity may occur. Under the assumption of a constant
parameter for the equation of state, we derive in this report a
rather general solution for the scale factor of a universe
dominated by tachyon matter, and show that, quite similarly to as
it happens in current quintessence models, that solution predicts
the occurrence of a big rip singularity. In the case that a
sub-quantum potential is added to the theory, we also obtain the
same result, though in this case the scale factor differs from the
expression obtained for current quintessence and "classical"
tachyon fields by terms which generally depend on the sub-quantum
potential, and the regime of phantom energy is restricted by the
smallness of the field kinetic term which may even make such a
regime to vanish.

The paper can be outlined as follows. In Sec. II we discuss a
rather general solution for the scale factor which satisfies all
requirements and constraints imposed by tachyon theory. The
phantom regime for such a solution is then investigated in Sec.
III, where it is seen that it shares all funny properties of
quintessential phantom energy. The sub-quantum generalization of
the tachyon theory is also considered in some detail in Sec. IV.
In this case it is shown that the accelerating expansion will
depend on the value of the sub-quantum potential, and that the
extend of the phantom regime inversely depends on the value taken
by the field kinetic energy. Results are summarized and briefly
discussed in Sec. V.

\section{Tachyon model for dark energy}

For the most favored cosmological spatially flat scenario, the
Friedmann equations read
\begin{equation}
H^2\equiv\left(\frac{\dot{a}}{a}\right)^2=\frac{8\pi G\rho}{3}
,\;\; \frac{\ddot{a}}{a}=-\frac{4\pi G(\rho+3P)}{3} ,
\end{equation}
where $\rho=\rho_{NR}+\rho_{R}+\rho_{\phi}$ is the energy density
for, respectively, non-relativistic, relativistic and tachyon
matter, and $P$ is the corresponding pressure. We shall restrict
ourselves to consider a description of the current cosmic
situation where it is assumed that the tachyon component largely
dominates and therefore we shall disregard in what follows the
non-relativistic and relativistic components of the matter density
and pressure. For the tachyon field $\phi$ we have [9]
\begin{equation}
\rho_{\phi}=\frac{V(\phi)}{\sqrt{1-\dot{\phi}^2}} ,\;\;
P_{\phi}=-V(\phi){\sqrt{1-\dot{\phi}^2}} ,
\end{equation}
in which $V(\phi)$ is the tachyon potential energy. Assuming an
equation of state $P_{\phi}=\omega_{\phi}\rho_{\phi}$ for the
tachyon matter, we then deduce that
\begin{equation}
\omega_{\phi}=\dot{\phi}^2-1 .
\end{equation}
Finally, the equation of motion for $\phi$ is
\begin{equation}
\ddot{\phi}+\left(1-\dot{\phi}^2\right)\left[3H\dot{\phi}
+\frac{1}{V(\phi)}\frac{dV(\phi)}{d\phi}\right] = 0.
\end{equation}

We shall show next that there exists a general solution for the
scale factor $a(t)$ in this tachyon-field scenario which has
exactly the same dependence on time as that in the general
solution for a pure quintessence scalar field, and hence we also
show that such a scenario admit the existence of a tachyon phantom
field which leads to a singularity in finite time. In fact, a
recipe has been provided by Padmanabhan himself [9] according to
which, given the explicit form for the scale factor $a(t)$, a
complete specification of the full $\phi$-field theory can be
achieved by using the following relations:
\begin{equation}
\frac{\dot{\rho}}{\rho}=2\frac{\dot{H}}{H}
\end{equation}
\begin{equation}
\dot{\phi}=\left(-\frac{2}{3}\frac{\dot{H}}{H^2}\right)^{1/2}
\end{equation}
\begin{equation}
V=\frac{3H^2}{8\pi
G}\left(1+\frac{2}{3}\frac{\dot{H}}{H^2}\right)^{1/2} .
\end{equation}
Our task then is to choose a general expression for $a(t)$ which
simultaneously satisfies relations (5), (6) and (7), together with
the Friedmann equations (1) and the equation of motion for the
tachyon field (4) which be able to match the accelerating
behaviour of the current universe and implies a physically
reasonable and suitably motivated field potential. If we assume a
linear time-dependence of the tachyon field $\phi$ and hence
constancy of parameter $\omega$, then it is not difficult to check
that a general form of such an expression for $a(t)$ can be
written as
\begin{equation}
a(t)=\left[a_0^{3\left(1+\omega_{\phi}\right)/2}+
\frac{3}{2}\left(1+
\omega_{\phi}\right)t\right]^{2/\left[3\left(1+
\omega_{\phi}\right)\right]} ,
\end{equation}
where $a_0$ is the initial value of the scale factor. We note that
this solution describes an accelerating universe in the interval
$-1/3>\omega_{\phi}>-1$. At the extreme point
$\omega_{\phi}=-1/3$, $a(t)$ describes a universe whose size
increases just as $t$, such as it should be expected. It is worth
realizing that by simply trivially re-scaling the time parameter,
solution (8) turns out to be nothing but the scale factor that
represents the most general solution for the case of a
quintessence scalar field for a constant equation of state [4]. On
the other hand, for a scale factor (8) the tachyon field and
potential are given by
\begin{equation}
\phi=\phi_0+\sqrt{1+\omega_{\phi}}\;t
\end{equation}
\begin{equation}
V(\phi)=\frac{3\sqrt{-\omega_{\phi}}}{8\pi
G\left[a_0^{3(1+\omega_{\phi})/2}+\frac{3}{2}\sqrt{1+
\omega_{\phi}}(\phi-\phi_0)\right]^2} .
\end{equation}
We note that as $\phi\rightarrow\infty$ this potential reasonably
vanishes after taking the form already considered by Padmanabhan
and others [9]. As $\phi\rightarrow\phi_0$ at $t=0$ $V(\phi)$
tends to a finite constant value, so clearly separating from the
unphysical behaviour of the potential considered by Padmanabhan
and compatible with what can be supported by string theories [9].
We regard therefore potential (11) as being physically reasonable.
Finally, we obtain for the speed of sound
\begin{equation}
c_s^2=\frac{\dot{P_{\phi}}}{\dot{\rho_{\phi}}}=\omega_{\phi} .
\end{equation}
For the accelerating-expansion regime, we see thus that the speed
of sound becomes imaginary, a case which could imply a collapsing
of the tachyon stuff that can still be circumvented however [12].

\section{Takyon phantom cosmology}

The phantom energy regime will be characterized by values of the
state equation parameter such that $\omega_{\phi}<-1$ and a
consequent violation of the dominant energy condition, i.e.
\begin{equation}
P{\phi}+\rho_{\phi}=\frac{V(\phi)\dot{\phi}^2}{\sqrt{1-
\dot{\phi}^2}}<0 .
\end{equation}
Such a regime can be obtained by simply Wick rotating the tachyon
field so that $\phi\rightarrow i\Phi$, with which the field $\Phi$
can be viewed as an axion tachyon field [13], as the scale factor
$a(t)$ and the field potential $V(\Phi)$ keep being positive and
given respectively by
\begin{equation}
a(t)=\left[a_0^{-3\left(|\omega_{\phi}|-1\right)/2}-
\frac{3}{2}\left(|\omega_{\phi}|-1\right)t\right]^{-2/\left[3\left(|\omega_{\phi}|
-1 \right)\right]} ,
\end{equation}
which accounts for a big rip singularity at finite future time
\begin{equation}
t_* =\frac{2}{3(|\omega_{\phi}|-1)a_0^{3(|\omega_{\phi}|-1)/2}} ,
\end{equation}
and
\begin{equation}
V(\Phi)=\frac{3\sqrt{|\omega_{\phi}|}}{8\pi
G\left[a_0^{-3(|\omega_{\phi}|-1)/2}-\frac{3}{2}\sqrt{|\omega_{\phi}|-
1}(\Phi-\Phi_0)\right]^2} ,
\end{equation}
with $\Phi_0\rightarrow -i\phi_0$. We note that both this
potential and the phantom tachyon energy density,
\begin{equation}
\rho_{\Phi}=\frac{3}{8\pi G\left[a_0^{-3(|\omega|-1)/2}
-\frac{3}{2}(|\omega|-1)t\right]^2} ,
\end{equation}
increase with time up to blowing up at $t=t_{*}$, to steadily
decrease toward zero thereafter. Thus, the tachyon model for dark
energy contains a regime for phantom energy which preserves all
the weird properties shown by this in current quintessence and
k-essence scenarios; i.e. superluminal speed of sound, increasing
energy density, violation of dominant energy condition and a big
rip singularity.

\section{Sub-quantum phantom cosmology}

If we extend next the concept of tachyon dark energy to include
also a sub-quantum potential $V_{SQ}$, then the Lagrangian for the
system can be generalized to read [11]:
\begin{equation}
L=-V(\phi)E(x,k),
\end{equation}
where $E(x,k)$ is the elliptic integral of the second kind [14],
with
\begin{equation}
x\equiv x(\phi)=\arcsin\sqrt{1-\dot{\phi}^2} ,\;\; k\equiv
k(\phi)=\sqrt{1-\frac{V_{SQ}^2}{V(\phi)^2}} ,
\end{equation}
in case that we consider a FRW spacetime. We note that in the
limit $V_{SQ}\rightarrow 0$, the above Lagrangian reduces to the
simple expression $L=-V(\phi)\sqrt{1-\dot{\phi}^2}$ which is the
Lagrangian of the tachyon theory discussed in Secs. II and III.
Defining as the pressure and energy density,
\begin{equation}
p_{\phi}=-V(\phi)E(x,k)
\end{equation}
\begin{equation}
\rho_{\phi}= \frac{V(\phi)\sqrt{1- \frac{\Delta
V^2(1-\dot{\phi}^2)}{V(\phi)^2}}\dot{\phi}}{\sqrt{1-
\dot{\phi}^2}}+V(\phi)E(x,k) ,
\end{equation}
where $\Delta V^2=V(\phi)^2-V_{SQ}^2$, we get for the sub-quantum
model again relation (5), together with
\begin{equation}
\frac{\sqrt{1-\frac{\Delta V^2(1-
\dot{\phi}^2)}{V(\phi)^2}}\dot{\phi}}{\sqrt{1-\frac{\Delta V^2(1-
\dot{\phi}^2)}{V(\phi)^2}}\dot{\phi}+E(x,k)\sqrt{1-\dot{\phi}^2}}
=-\frac{2\dot{H}}{3H^2} .
\end{equation}
and
\begin{equation}
V(t)=-\left[\left(\frac{2\dot{H}}{8\pi G}\right)^2 -\dot{\phi}^2
V_{SQ}^2\right]^{1/2}\frac{\sqrt{1-\dot{\phi}^2}}{\dot{\phi}^2} .
\end{equation}
Now, if an expression like (8) is taken for the scale factor, then
$V(t)$ would not vanish in the limit that $t\rightarrow\infty$
(actually $V$ becomes an imaginary finite constant at that limit),
a behaviour which cannot be accepted as reasonable. Therefore, a
form as that is given by Eq. (8) appears here as being not
suitable for the scale factor. A Hubble parameter which satisfies
however the requirement that $V(t)\rightarrow 0$ as
$t\rightarrow\infty$ is
\begin{equation}
H=-\frac{1}{6(1+\omega)}\left\{\sigma(\phi,V_{SQ},t) +2\pi
G\dot{\phi}V_{SQ}\ln\left[\frac{\sigma(\phi,V_{SQ},t) -4\pi
G\dot{\phi}V_{SQ}}{\sigma(\phi,V_{SQ},t) +4\pi
G\dot{\phi}V_{SQ}}\right]\right\} ,
\end{equation}

with
\begin{equation}
\sigma(\phi,V_{SQ},t)=\sqrt{(4\pi G\dot{\phi}V_{SQ})^2
+\frac{9(1+\omega)^2}{\left(a_0^{3(1+\omega)/2}
+\frac{3}{2}(1+\omega)t\right)^4}} .
\end{equation}
Even though we have not been able to integrate Eq. (23) so that a
closed-form expression for the scale factor could be obtained, we
are already prepared to check whether or not the tachyon theory
equipped with a sub-quantum potential can show a big rip
singularity.

From Eqs. (19) and (20) we can in general obtain
\begin{equation}
\omega=-\frac{E(x,k)V(\phi)\sqrt{1- \dot{\phi}^2}}{\sqrt{\Delta
V^2\dot{\phi}^2+ V_{SQ}^2}\dot{\phi} +
E(x,k)V(\phi)\sqrt{1-\dot{\phi}^2}}
\end{equation}
\begin{equation}
p_{\phi}+\rho_{\phi}=V(\phi)\frac{\sqrt{\Delta V^2\dot{\phi}^2+
V_{SQ}^2}\dot{\phi}}{V(\phi)\sqrt{1-\dot{\phi}^2}} .
\end{equation}
It follows from these two expressions that a regime with phantom
behaviour showing weird properties similar to those found in the
phantom tachyon model can also be achieved by Wick rotating
$\phi\rightarrow i\Phi$, if and only if
\[\Delta V^2 \dot{\Phi}^2>V_{SQ}^2 \]
and
\[\sqrt{\Delta V^2\dot{\Phi}^2-V_{SQ}^2}\dot{\Phi}
<V(\Phi)E(x,k)\sqrt{1+\dot{\Phi}^2} ,\] in which now $\Delta
V^2=V(\Phi)^2-V_{SQ}^2$, $x=\sinh^{-1}\sqrt{1+\dot{\Phi}^2}$, and
$k=\sqrt{1-\frac{V_{SQ}^2}{V(\Phi)^2}}$. Note that such a regime
will only be possible for reasonably large values of
$|\dot{\Phi}|$.

On the other hand, by closely inspecting Eqs. (23) and (24) it is
not difficult to convince oneself that in the regime where
$\omega<-1$ both the scale factor
\[a=\exp\left(\int Hdt\right) \]
(which in fact describes an accelerating universe for $\omega>-1$)
and the potential (22) blow up at the finite time $t_*$ if
$\omega<-1$, so indicating the presence of the big rip singularity
in the sub-quantum phantom domain. We have thus shown that the
presence of phantom energy in the sub-quantum cosmological model
also leads to a big rip doomsday.

\section{Conclusions and further comments}

By using tachyon-like theories whose starting Lagrangians
generally are inspired by string theories, we have investigated
the properties of general FRW cosmological solutions that are
fully consistent with the whole dynamical structure of the
theories. Assuming a general equation of state $p=\omega\rho$,
such solutions describe accelerating universes in the interval
$-1/3>\omega>-1$ and, among other weird properties characterizing
the phantom regime, all show a big rip singularity in finite time
when $\omega<-1$. If in deriving the Lagrangian we begin with a
relativistic mechanical momentum-energy quantum relation
interpretable in terms of a sub-quantum potential, then the
existence of a phantom regime will depend on how much
slowly-varying the scalar field is allowed to be. Thus, whenever
the kinetic term for the field is smaller or comparable with the
sub-quantum potential then the phantom regime cannot exist in this
kind of tachyon theory.

A potential problem with the tachyon models considered in this
paper stems from the imaginary value of the sound speed. In fact a
value $c_s^2<0$ implies occurrence of instability on scales below
the Jeans limit for scalar field fluctuations [15] which grow
therefore exponentially. Whereas when these models are considered
as pure cosmic vacuum components this could in fact be regarded as
an actual problem, if the tachyon models are viewed as the sum of
two components [9], one describing the negative pressure vacuum
stuff and the other describing the dust-like cold dark matter
contributing $\Omega_m\sim 0.35$ and clustering gravitationally at
small scales, the occurrence of instabilities arising from a
negative value for $c_s^2$ might instead be regarded as a way to
explain some recent observations in galaxies and superclusters
that concern gravitationally collapsed objects such as
supermassive black holes and dark-matter galactic halos. That
possibility deserves further consideration and is in fact being
the subject of a more thorough research by the author.

\noindent {\bf acknowledgements}

\noindent The author thanks Carmen L. Sig\"{u}enza for useful
discussions and Yun-Song Piao and Fabio Finelli for valuable
correspondence. This work was supported by DGICYT under Research
Project BMF2002-03758.

\noindent\section*{References}

\begin{description}

\item {1} A.C. Baccigalupi, A. Balbi, S. Matarrase, F. Perrotta
and N. Vittorio, Phys. Rev. D65, 063520 (2002); M. Melchiorri, L.
Mersini, C.J. Odman and M. Tradden, Phys. Rev. D68, 043509 (2003);
M. Doupis, A. Riazuelo, Y. Zolnierowski and A. Blanchard, Astron.
Astrophys. 405, 409 (2003); L. Tonry {\it el al.}, Astrophys. J.
594, 1 (2003); J.S. Alcaniz, astro-ph/0312424 . \item {2}R.R.
Caldwell, Phys. Lett. B545, 23 (2002). \item {3} B. McInnes, JHEP
0208, 029 (2002); G.W. Gibbons, hep-th/0302199; A.E. Schulz and
M.J. White, Phys. Rev. D64, 043514 (2001); J.G. Hao and X. Z. Li,
Phys. Rev. D67, 107303 (2003); S. Nojiri and S.D. Odintsov, Phys.
Lett.B562, 147 (2003); B565, 1 (2003); B571, 1 (2003); P.Singh, M.
Sami and N. Dadhich, Phys. Rev. D68, 023522 (2003); J.G. Hao and
X.Z. Li, Phys. Rev. D68, 043501; 083514 (2003); X.Z. Li and J.G.
Hao, hep-th/0303093, Phys. Rev. D (in press); M.P. Dabrowski, T.
Stachowiak and M. Szydlowski, Phys. Rev. D68, 067301 (2003); M.
Szydlowski, W. Zaja and A. Krawiec, astro-ph/0401293; E. Elizalde
and J. Quiroga H., Mod. Phys. Lett. A19, 29 (2004); V.B. Johri,
astro-ph/0311293; L.P. Chimento and R. Lazkoz, Phys. Rev. Lett.
91, 211301 (2003); H.Q. Lu, hep-th/0312082; M. Sami, A.
Toporensky, gr-qc/0312009; R. Naboulsi, gr-qc/0303007, Class.
Quan. Grav (in press); X.H. Meng and P. Wang, hep-ph/0311070; H.
Stefancic, astro-ph/0310904, Phys. Lett. B (in press) ;
astro-ph/0312484; D.J. Liu and X.Z. Li, Phys. Rev. D68, 067301
(2003); A. Yurov, astro-ph/0305019; Y.S. Piao and E. Zhou, Phys.
Rev. D68, 083515 (2003);Y.S. Piao and Y.Z. Zhang,
astro-ph/0401231; H.Q. Lu, hep-th/0312082 ; M. Szydlowski, W.
Czaja and A. Krawiec, astro-ph/0401293; J.M. Aguirregabiria, L.P.
Cimento and R. Lazkoz, gr-qc/0403157; J. Cepa, astro-ph/0403616,
Astron. Astrophys. (in press); V. Faraoni, gr-qc/0404078, Phys.
Rev. D. (in press); Z.-K. Guo, Y.-S. Piao and Y.-Z. Zhang,
astro-ph/0404154; E. Elizalde, S. Nojiri and S. Odintsov,
hep-th/0405034; Y.-H. Wei and Y. Tian, gr-qc/0405038; F. Piazza
and S. Tsujikawa, hep-th/0405054; S. Nojiri and S.D. Odinsov,
hep-th/0405078. \item {5} R.R. Caldwell, M. Kamionkowski and N.N.
Weinberg, Phys. Rev. Lett. 91, 071301 (2003); J.D. Barrow, Class.
Quant. Grav. 21, L79 (2004). \item {6} J.D. Barrow, G.J. Galloway
and F.J. Tipler, Mon. Not. R. Astron. Soc. 223, 835 (1986). \item
{4} P.F. Gonz\'{a}lez-D\'{\i}az, Phys. Rev. D68, 084016 (2003). \item {5}
R.R. Caldwell, M. Kamionkowski and N.N. Weinberg, Phys. Rev. Lett.
91, 071301 (2003). \item {6} P.F. Gonz\'{a}lez-D\'{\i}az, Phys. Lett. B586,
1 (2004). \item {7} A. Kamenshchik, U. Moschella, V. Pasquier,
Phys. Lett. B511, 265 (2001); N. Bilic, G.B. Tupper and R.
Viollier, Phys. Lett. B535, 17 (2001); M.C. Bento, O. Bertolami
and A.A. Sen, Phys. Rev. D66, 043507 \item {8} P.F. Gonz\'{a}lez-D\'{\i}az,
Phys. Rev. D68, 021303(R) (2003); M. Bouhmadi and J.A.
Jim\'{e}nez-Madrid, astro-ph/0404540. \item {9} T. Padmanabhan, Phys.
Rev. D66, 021301 (2002); T. Padmanabhan and T.R. Choudhury, Phys.
Rev. D66, 081301 (2002); J.S. Bagla, H.K. Jassal, T. Padmanabhan,
Phys. Rev. D67, 063504 (2003) \item {10} P.F. Gonz\'{a}lez-D\'{\i}az, Phys.
Rev. D69, 103512 (2004) . \item {11} L.R.W. Abramo and F. Finelli,
Phys. Lett. B575, 165 (2003). \item {12} T. Padmanabhan, {\it
Structure Formation in the Universe} (Cambridge University Press,
Cambridge, UK, 1993). \item {13} P.F. Gonz\'{a}lez-D\'{\i}az, Phys. Rev.
D69, 063522 (2004) . Report-no. IMAFF-RCA-03-08 \item {14} M.
Abramowitz and I.A. Stegun, {\it Handbook of Mathematical
Functions} (Dover, New York, USA, 1965). \item {15} D. Carturan
and F. Finelli, Phys. Rev. D68, 103501 (2003); H.B. Sandvik, M.
Tegmark, M. Zaldarriaga and I. Waga, astro-ph/0212114 .

\end{description}

\end{document}